# High-resolution analysis of the $\nu_3$ band of radiocarbon methane $^{14}CH_4$

Santeri Larnimaa[1], Lauri Halonen[1], Juho Karhu[1], Teemu Tomberg[1], Markus Metsälä[1], Guillaume Genoud[2], Tuomas Hieta[3], Steven Bell[4], Markku Vainio[1,5]

1. Department of Chemistry, University of Helsinki, Helsinki, Finland
2. VTT Technical Research Centre of Finland Limited, Espoo, Finland
3. Gasera Ltd., Turku, Finland
4. National Physical Laboratory, Middlesex, UK
5. Photonics Laboratory, Physics Unit, Tampere University, Tampere, Finland

**Abstract:** We present the first high-resolution measurements and rotational analysis of the fundamental asymmetric stretching vibrational band $\nu_3(F_2)$ of radiocarbon methane. A spectrometer consisting of a mid-infrared continuous-wave optical parametric oscillator and a cantilever-enhanced photoacoustic detector was employed to determine the wavenumbers of 43 rovibrational lines. A spectroscopic model could reproduce all the observed transition wavenumbers within the accuracy of the experiment. Our work contributes to the development of radiocarbon methane detectors based on laser spectroscopy.

**Introduction**

Optical spectroscopy of the radioactive isotopologue of methane ($^{14}CH_4$) has interesting applications ranging from energy industry to atmospheric sciences. As an example, radioactive gaseous emissions of light water nuclear reactors and their decommissioning sites can be monitored by measuring the $^{14}CH_4$ concentration [1, 2]. Since the $^{14}CH_4$ concentration is depleted in fossil sources with a $^{14}C$ half-life of 5700 years, measurements of the $^{14}CH_4/^{12}CH_4$ ratio make it possible to determine the biofractions of biogas and natural gas mixtures, as well as to apportion methane emission sources [3, 4]. Additionally, the estimates of fossil fraction of the global total $CH_4$ emissions are mainly based on observations of atmospheric radiocarbon methane [5, 6].

Accelerator mass spectrometry (AMS) and liquid scintillation counting are currently the standard methods for measurements of small concentrations of long-lived radioactive isotopes. They can deliver extremely high sensitivity for $^{14}C$ measurements, but they often require complex sample preparation and are unsuitable for in-situ measurements. In contrast, optical laser spectroscopy is inherently suitable for the development of compact, automated, and cost-effective instruments that are ideal for online in-situ monitoring. Optical methods are also species-selective, the required measurement time is independent of radioactive decay rate, and the radiocarbon content can easily be determined even if the sample contains other radionuclides. This is not the case with, for example, liquid scintillation counting.

The feasibility of optical spectroscopic detection of radioactive gas-phase species in low concentrations was investigated already in the early 1980s [7], but actual measurements were demonstrated much later using highly sensitive cavity-enhanced techniques. Laser spectroscopy of radioactive carbon compounds has so far concentrated on radiocarbon dioxide [2, 8-13]. An optical $^{14}CO_2$ detector with sensitivity approaching that of AMS has already been demonstrated [14]. The development of similar instruments for $^{14}CH_4$ has been hindered by the lack of fundamental spectroscopic information. The first reported infrared spectrum of radiocarbon methane was published only recently: we recorded the $\nu_3(F_2)$ asymmetric stretching fundamental band using a new spectroscopic method that combines broadband photoacoustic detection with an optical frequency comb [15].

In this letter, we present an absorption spectrum of the $\nu_3(F_2)$ band of $^{14}CH_4$ recorded in higher resolution and with much better sensitivity than in our previous work [15]. The experimental improvements make it possible to observe several $^{14}CH_4$ lines that were previously too weak to be measured or were concealed by absorption lines of normal methane ($^{12}CH_4$). The new and improved data allow us to analyze the rotational structure of $^{14}CH_4$ using a more advanced spectroscopic model than in Ref. [15]. The obtained accurate line position information benefits the development of optical radiocarbon methane detectors.

The $^{14}CH_4$ absorption lines observed in this work are summarized in Fig. 1, which shows the measured line positions with respect to the $\nu_3(F_2)$ band of the $^{12}CH_4$ isotopologue. The measurements were carried out using cantilever enhanced photoacoustic spectroscopy (CEPAS), which combines high detection sensitivity with a small sample volume, thus being ideal for samples with limited availability or concentration [16]. Since the photoacoustic signal is directly proportional to the absorbed optical power, the detection sensitivity of CEPAS can be maximized by using a high-power mid-infrared light source, such as a narrow-linewidth optical parametric oscillator (CW OPO) [17, 18]. The same approach was employed in this study. The experimental setup is described in the next section, after which we report the first high-resolution measurements and rotational analysis of radiocarbon methane.

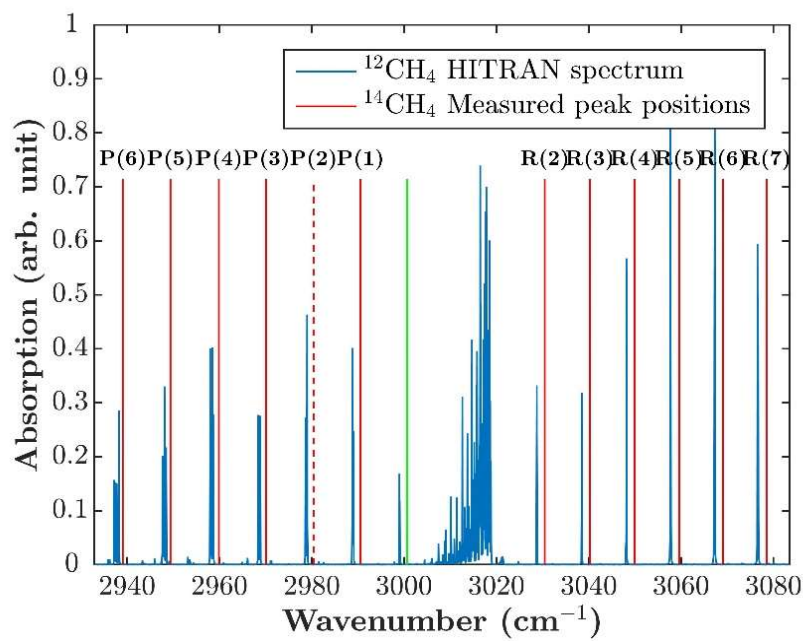

*Fig. 1. The red vertical lines indicate the positions of measured $^{14}CH_4$ absorption lines, and the green vertical line denotes the band centre. The blue spectrum is a HITRAN simulation of the $^{12}CH_4$ spectrum. The band centre of $^{14}CH_4$ is shifted by about −21 $cm^{-1}$ relative to that of $^{12}CH_4$. The peaks have tetrahedral fine structure, which makes the total number of measured $^{14}CH_4$ lines 43. The bolded labels denote lines that were also measured in our earlier work [15]. The P(2) line is indicated by a dashed line, because its fine structure could not be reliably determined due to significant spectral interference caused by water impurity in the gas sample.*

**Experimental methods**

The main building blocks of the experimental setup are shown in Fig. 2. The gas sample contains 1 ppm of $^{14}CH_4$ and 100 ppm of $^{12}CH_4$ mixed in nitrogen (Quotient Bioresearch). Prior to each measurement, the gas line and the photoacoustic cell were flushed with pure nitrogen for several seconds, and subsequently evacuated using a vacuum pump. The system was then flushed with the sample gas and pumped to the final measurement pressure (20 kPa). The cell inlet and outlet valves were closed during the measurements in order to have a static gas sample in the photoacoustic cell. Despite these precautions, minor but detectable traces of water were present as impurity in the photoacoustic cell due to small leaks and outgassing.

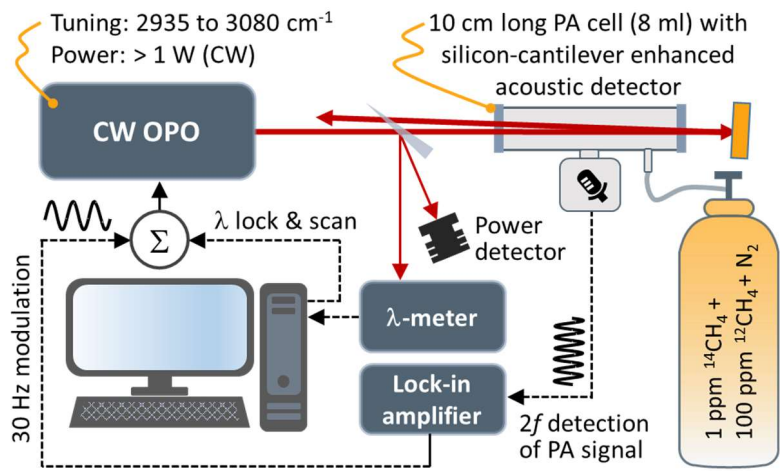

***Fig. 2.*** *Experimental setup. Optical and electrical signals are shown by solid and dashed arrows, respectively. The CW OPO output beam is double-passed through the photoacoustic (PA) gas cell, which is equipped with a sensitive acoustic detector. Small fractions of the CW OPO beam are sampled in order to precisely scan the mid-infrared frequency and monitor the mid-infrared power. The PA signal is produced by wavelength-modulation spectroscopy and measured with a lock-in amplifier. The voltage signal required for sinusoidal CW OPO wavelength modulation is produced by the lock-in amplifier and summed with the lock-and-scan signal prior to sending it to the CW OPO.*

The CW OPO light source is similar to other singly-resonant mid-infrared CW OPOs developed in our laboratory [17, 19]. It is tuneable from 2935 to 3080 cm$^{-1}$ and covers a large fraction of the $\nu_3(F_2)$ band of $^{14}CH_4$, as illustrated in Fig. 1. The mid-infrared output power of the CW OPO exceeds 1 W within the entire tuning range, making it possible to reach ppb-level detection sensitivity for methane when combined with CEPAS. High spectral resolution is guaranteed by the narrow linewidth, *c.a.* 1 MHz, and active stabilization of the CW OPO. The mid-infrared frequency was locked to a wavelength meter (Bristol 771B) using a computer-controlled feedback loop, which regulated the mid-infrared frequency via a piezoelectric actuator that controls the CW OPO pump laser frequency [18]. Automated high-precision scans were thus possible by adjusting the lock point. The frequency axis was calibrated using $^{12}CH_4$ and water peaks that fell within the measurement range and are listed in the HITRAN database [20]. With $\tilde{\nu}_{meas}$ being the measured peak position and $\tilde{\nu}_{HITRAN}$ being the respective HITRAN value, the wavenumber difference $\Delta\tilde{\nu}_{cal} = \tilde{\nu}_{meas} - \tilde{\nu}_{HITRAN}$ was determined for each of the 28 calibration peaks. The mean of all $\Delta\tilde{\nu}_{cal}$ revealed a wavenumber offset of –0.007 cm$^{-1}$, which was corrected. The calculated standard deviation of all $\Delta\tilde{\nu}_{cal}$ is 0.0013 cm$^{-1}$.

The CW OPO output beam was directed to the CEPAS cell (Gasera PA201, volume 8 ml), which contained the gas sample at room temperature. The gas pressure in the cell was set to 20 kPa in order to reduce pressure broadening and thus to better resolve the closely packed tetrahedral components of the $^{14}CH_4$ absorption lines. Further improvements of spectral selectivity and signal-to-noise ratio (SNR) were obtained by applying wavelength modulation spectroscopy with 2*f* detection: The CW OPO mid-infrared wavelength was dithered with a sinusoidal signal at the modulation frequency *f* = 30 Hz and the photoacoustic signal was demodulated at 2*f* = 60 Hz using a lock-in amplifier. The voltage signal for modulation was combined with the scan voltage in an electronic summation circuit before passing it on to the piezoelectric actuator of the pump laser. Both the frequency and amplitude of wavelength modulation were empirically optimized for CEPAS detection [17]. In order to compensate for changes in the CW OPO output power during wavelength scans, the demodulated photoacoustic signal was normalized by the mid-infrared power incident on the CEPAS cell. The power was continuously monitored with a mid-infrared photodetector, in order to compensate for CW OPO power fluctuations that can be up to a few percent during wavelength scanning.

**Results**

Having prior information on the $^{14}CH_4$ line positions [15], we did not have to perform high-resolution scans over the entire spectral range shown in Fig. 1. Instead, we only scanned across the expected line positions. An example of a measured high-resolution spectrum for line $R$(3) is illustrated in Fig. 3. The individual tetrahedral components are easier to identify in the lower panel of the figure, which shows the reconstructed absorption spectrum. The spectrum was reconstructed by fitting a simulated 2$f$ spectrum to the measured 2$f$ spectrum assuming Voigt line profiles. The modulation amplitude was fixed to a separately measured value (243 MHz), while the pressure broadening was a free parameter in the fit. The Doppler widths were fixed to their calculated values. In addition to the fitted $A_1$, $F_1$ and $F_2$ line components of radiocarbon methane, the simulation of Fig. 3 includes weak overlapping lines of $^{12}CH_4$. The overlapping lines were either fitted or computed using the respective HITRAN values, depending on the line strengths. The same procedure was used to account for water peaks that in some cases interfered with the radiocarbon spectrum.

The measurement SNR for the example of Fig. 3 exceeds 400, which corresponds to a noise-equivalent detection limit below 3 ppb. This is a factor of 30 improvement compared to the previous measurement of the same line [15]. The spectral resolution of the CW OPO measurements reported here is an order of magnitude better than what was possible to obtain in our earlier work, where the resolution (0.02 cm$^{-1}$) was limited by the Fourier-transform infrared spectrometer used in the measurements. The considerable improvements in both SNR and spectral resolution offered by the CW OPO-based setup allowed us to reliably measure in total 43 line components, 17 of which are new, previously unobserved components. In particular, this time we were able to measure accurately several lines of the $P$ branch, which was difficult to address in the first experiment. The only exception is line $P$(2), which was also excluded in the present study because it severely overlaps with strong water absorption, thus hindering reliable determination of the line components. All other measured line positions are listed in Table 2. The estimated uncertainty of the determined line positions is smaller than 0.003 cm$^{-1}$ ($k$ = 2) for all lines reported in the table. The combined uncertainty is dominated by the uncertainty of the abovementioned wavenumber-axis calibration, while the fit uncertainties are typically of the order of 3×10$^{-4}$ cm$^{-1}$.

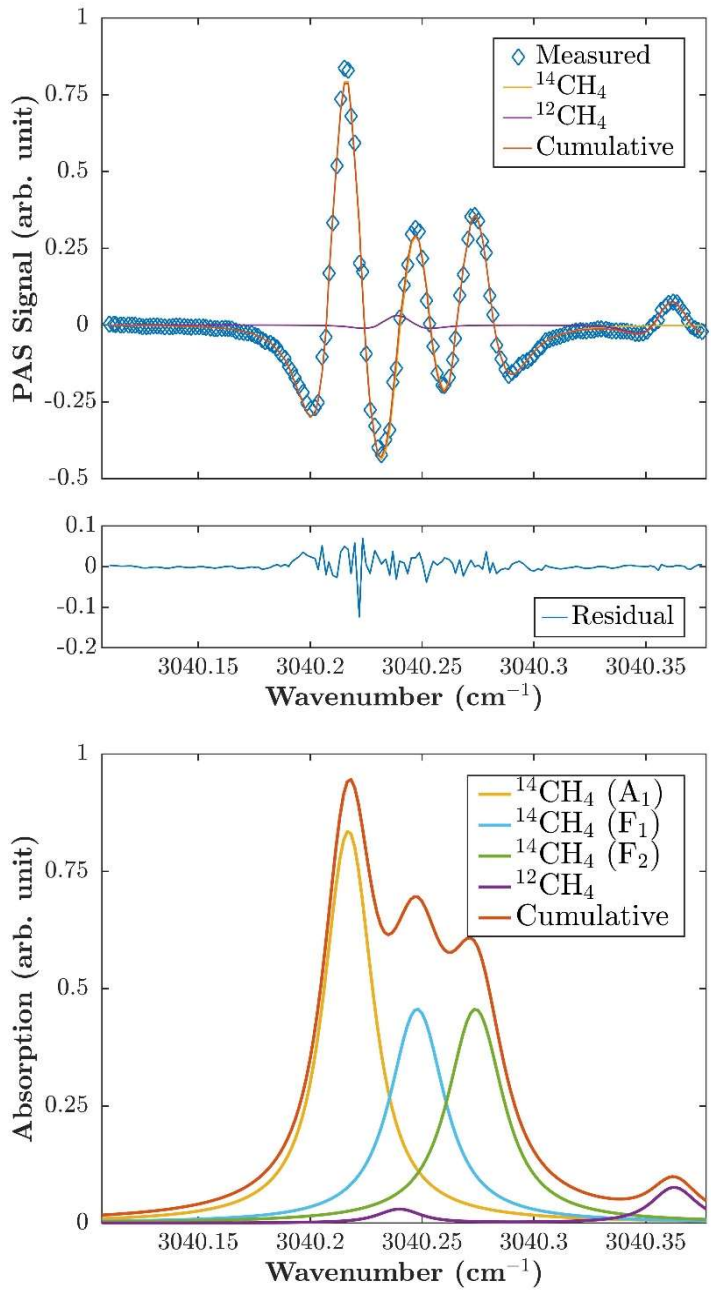

*Fig. 3*. Upper panel: The 2f CEPAS signal as a function of wavenumber for the line R(3) of the $\nu_3$ band of $^{14}CH_4$. Also the fit residual is shown. Lower panel: Absorption spectrum reconstructed from the measured 2f CEPAS spectrum.

**The rotational analysis of the $\nu_3(F_2)$ band of $^{14}CH_4$**

Our rotational analysis of the $\nu_3(F_2)$ fundamental band of $^{14}CH_4$ is based on an isolated band model, which has previously been used to assign the same band of normal methane [21]. It is well-known from the prior work done with $^{12}CH_4$ that the upper states of the $\nu_3(F_2)$ band are coupled via Fermi, Coriolis and α-resonances with the upper states of bending overtones and combination bands, as well as with the $\nu_1(A_1)$ band [21]. However, we did not use the more sophisticated approaches [22] as our spectral data are limited to small values of the total angular momentum quantum number *J*.

The Hamiltonians we use are for the ground vibrational state [23, 24]

$$H_0 = B_0 J(J+1) - D_0 J^2(J+1)^2 + H_0 J^3(J+1)^3 - \left[D_t^{(0)} + H_{4t}^{(0)} J(J+1)\right] \mathbf{T}_{044} + H_{6t}^{(0)} \mathbf{T}_{066}$$

and for the F$_2$ state [24]

$$H_{\nu_3} = \nu_3 + B_{\nu_3} J(J+1) - D_{\nu_3} J^2(J+1)^2 + H_{\nu_3} J^3(J+1)^3 - 2B\zeta_3 \mathbf{T}_{110} + \alpha_{220}\mathbf{T}_{220} + \alpha_{224}\mathbf{T}_{224} - \left[D_t^{(\nu_3)} + H_{4t}^{(\nu_3)} J(J+1)\right] \mathbf{T}_{044} + H_{6t}^{(\nu_3)} \mathbf{T}_{066}$$

where $\nu_3$ is the band origin, $B$ is the rotational constant, $\zeta_3$ is the Coriolis constant, and $D$ and $H$ are the quartic and sextic centrifugal distortion constants, respectively. The $B$, $D$, and $H$ constants are scalar ones and $D_t$, $H_{4t}$, $H_{6t}$ are coefficients of tensor terms. The quantity $\mathbf{T}_{ijk}$ is a rank $k$ spherical tensor operator consisting of vibrational operators of the rank $i$ and rotational operators of the rank $j$.

The Hamiltonian matrices for the lower and upper state $J$ values were set up using products of harmonic oscillator and rigid symmetric rotor eigenfunctions. The uncoupled basis was adopted for the F$_2$ vibrational state. The computer program employed here was originally written for the ro-vibrationally coupled $\nu_1$(A$_1$)/$\nu_3$(F$_2$) problem of SiH$_4$ [25]. We excluded ro-vibrational coupling between the A$_1$ and F$_2$ states of $^{14}$CH$_4$.

Our experimental spectrum was first preliminarily assigned using the similarity with the $\nu_3$(F$_2$) band transitions of $^{12}$CH$_4$ [15, 20]. We used the spectroscopic model described above to confirm these assignments. The non-linear least squares method was employed to optimize the upper state ro-vibrational parameter values as given in Table 1. All lower state parameters and the upper state sextic centrifugal distortion and $\alpha_{220}$ coefficients have been constrained to the $^{12}$CH$_4$ and $^{13}$CH$_4$ values as given in Table 1 [21, 26]. All observed and computed transitions can be found in Table 2. The standard deviation of the least squares fit is 0.003 cm$^{-1}$, thus confirming the high internal accuracy of the measured transition wavenumbers.

The data at our disposal do not support optimization of the ground state parameters. Fortunately, the similarity between the ground state rotational constants and between the ground state centrifugal distortion constants of $^{12}$CH$_4$ and $^{13}$CH$_4$ supports the constraints we have made [26]. All upper state coefficients we have optimized are well determined and are similar to the isolated $\nu_3$(F$_2$) band model results of $^{12}$CH$_4$ [21]. We could not optimize the $\alpha_{220}$ parameter, which is important in the local mode theory [27-29].

The last column of Table 2 lists the computed relative intensities of the $^{14}$CH$_4$ lines studied here. This information is included as a reference for future experimental work that is needed to accurately determine the line strengths. While the wavelength-modulation CEPAS method used here is excellent for line position measurements, it is not ideal for accurate intensity measurements. Absolute line strength measurements could be done using cavity ring-down spectroscopy [11], which among other applications has previously been applied to accurately determine the line parameters of the $\nu_3$ band of $^{14}$CO$_2$ [13].

**Table 1.** Spectroscopic parameters of $^{14}CH_4$. The statistical uncertainties given in parentheses are one-standard errors in the least significant digits.

| Parameter (cm$^{-1}$) | Ground state | $v_3(F_2)$ |
|---|---|---|
| $v$ | | 3001.05438 (140) |
| $B$ | 5.241288289[a] | 5.2010152 (880) |
| $D/10^{-4}$ | 1.10876[b] | 0.8662 (110) |
| $H/10^{-9}$ | 5.77[b] | 5.77[b] |
| $D_t/10^{-6}$ | 4.434515[b] | 6.562 (170) |
| $H_{4t}/10^{-10}$ | -5.6652[b] | -5.6652[b] |
| $H_{6t}/10^{-10}$ | 3.68[b] | 3.68[b] |
| $2B\zeta_3$ | | 0.394920 (120) |
| $\alpha_{220}$ | | -0.00827[b] |
| $\alpha_{224}$ | | -0.0023128 (130) |

[a]Constrained to the value for $^{13}CH_4$ from Ref. [26]
[b]Constrained to the value for $^{12}CH_4$ from Ref. [21]

**Table 2**. Observed transition wavenumbers $\tilde{\nu}_{obs}$ of the $\nu_3(F_2)$ band of $^{14}CH_4$, as well as the differences ($\Delta_{o-c}$) between the observed and calculated wavenumbers. The symmetry labels are given for both the upper ($\Gamma'$) and lower ($\Gamma''$) state. The ordering numbers of both the upper state ($n'$) and lower state ($n''$) increase in going down in energy within a given symmetry block. The last column gives the calculated relative intensities.

| Transition | $\Gamma'$ | $n'$ | $\Gamma''$ | $n''$ | $\tilde{\nu}_{obs}$ (cm$^{-1}$) | $\Delta_{o-c}$ (cm$^{-1}$) | Intensity |
|---|---|---|---|---|---|---|---|
| P(6) | $A_2$ | 1 | $A_1$ | 1 | 2938.858 | 0.003 | 57.0 |
| | $F_2$ | 1 | $F_1$ | 1 | 2938.938 | 0.003 | 33.9 |
| | $F_1$ | 2 | $F_2$ | 1 | 2939.023 | 0.001 | 33.6 |
| | $A_1$ | 1 | $A_2$ | 1 | 2939.246 | -0.001 | 57.3 |
| | $F_1$ | 1 | $F_2$ | 2 | 2939.333 | -0.002 | 34.1 |
| | E | 1 | E | 1 | 2939.356 | -0.003 | 22.9 |
| P(5) | $F_2$ | 2 | $F_1$ | 1 | 2949.366 | 0.000 | 37.7 |
| | E | 1 | E | 1 | 2949.419 | 0.000 | 25.0 |
| | $F_1$ | 1 | $F_2$ | 1 | 2949.612 | 0.001 | 38.0 |
| | $F_2$ | 1 | $F_1$ | 2 | 2949.700 | 0.001 | 38.0 |
| P(4) | $F_1$ | 1 | $F_2$ | 1 | 2959.755 | -0.004 | 37.8 |
| | E | 1 | E | 1 | 2959.901 | 0.000 | 25.4 |
| | $F_2$ | 1 | $F_1$ | 1 | 2959.955 | 0.002 | 38.0 |
| | $A_2$ | 1 | $A_1$ | 1 | 2960.022 | 0.003 | 63.5 |
| P(3) | $A_1$ | 1 | $A_2$ | 1 | 2970.058 | -0.006 | 55.6 |
| | $F_1$ | 1 | $F_2$ | 1 | 2970.160 | -0.001 | 33.2 |
| | $F_2$ | 1 | $F_1$ | 1 | 2970.235 | 0.003 | 33.3 |
| P(1) | $F_2$ | 1 | $F_1$ | 1 | 2990.573 | 0.001 | 8.6 |
| R(2) | $F_1$ | 3 | $F_2$ | 1 | 3030.474 | 0.001 | 54.2 |
| | E | 2 | E | 1 | 3030.487 | 0.002 | 36.2 |
| R(3) | $A_1$ | 1 | $A_2$ | 1 | 3040.217 | -0.002 | 100.0 |
| | $F_1$ | 3 | $F_2$ | 1 | 3040.248 | 0.000 | 59.8 |
| | $F_2$ | 4 | $F_1$ | 1 | 3040.274 | 0.001 | 59.8 |
| R(4) | $F_1$ | 4 | $F_2$ | 1 | 3049.898 | -0.002 | 59.8 |
| | E | 3 | E | 1 | 3049.952 | 0.001 | 39.5 |
| | $F_2$ | 4 | $F_1$ | 1 | 3049.974 | 0.001 | 59.4 |
| | $A_2$ | 2 | $A_1$ | 1 | 3050.007 | 0.002 | 99.3 |
| R(5) | $F_2$ | 5 | $F_1$ | 1 | 3059.487 | -0.001 | 54.8 |
| | E | 3 | E | 1 | 3059.503 | -0.001 | 36.5 |
| | $F_1$ | 5 | $F_2$ | 1 | 3059.586 | 0.003 | 53.9 |
| | $F_2$ | 4 | $F_1$ | 2 | 3059.629 | 0.000 | 54.3 |
| R(6) | $A_2$ | 2 | $A_1$ | 1 | 3068.972 | -0.003 | 78.0 |
| | $F_2$ | 5 | $F_1$ | 1 | 3068.996 | -0.001 | 46.6 |
| | $F_1$ | 6 | $F_2$ | 1 | 3069.020 | -0.004 | 46.3 |
| | $A_1$ | 2 | $A_2$ | 1 | 3069.126 | 0.007 | 75.4 |
| | $F_1$ | 5 | $F_2$ | 2 | 3069.166 | -0.001 | 45.5 |
| | E | 4 | E | 1 | 3069.180 | -0.002 | 30.5 |
| R(7) | $F_2$ | 7 | $F_1$ | 1 | 3078.397 | 0.001 | 37.1 |
| | E | 4 | E | 1 | 3078.431 | 0.005 | 24.5 |
| | $F_1$ | 6 | $F_2$ | 1 | 3078.453 | 0.002 | 36.4 |
| | $A_1$ | 2 | $A_2$ | 1 | 3078.501 | -0.008 | 61.5 |
| | $F_1$ | 5 | $F_2$ | 2 | 3078.631 | 0.004 | 35.5 |
| | $F_2$ | 6 | $F_1$ | 2 | 3078.652 | -0.002 | 35.9 |

## Conclusions

We have reported the first high-resolution measurements and analysis of the $v_3(F_2)$ fundamental stretching vibrational band of radiocarbon methane. Our new measurements have significantly better SNR and spectral resolution than the first measurements reported in Ref. [15]. As a result, we have observed several new line components and reduced the uncertainty of line position determination compared to the previous work. The new high-quality data enabled proper analysis of the ro-vibrational band, in contrast to an earlier simpler model that did not include *e.g.* a proper analysis of tetrahedral splittings and centrifugal distortion. While we have been able to measure accurately the positions of several lines of the *P* and *R* branches, the *Q* branch was obscured by strong overlapping spectral features of the stable isotopologue $^{12}CH_4$, which has 100 times higher concentration than $^{14}CH_4$ in the gas sample that was available for the experiments reported here. On the other hand, the spectroscopic model reported here can be used to precisely predict the yet unobserved line positions, including those of the *Q* branch. Together with the known absorption line positions of $^{12}CH_4$, $H_2O$ and other potential interfering gas species, the work published here makes it possible to identify the absorption lines that are suitable for precise and selective in-situ monitoring of radiocarbon methane.

## Acknowledgements

Financial support of the Academy of Finland is acknowledged (Decisions 314363 and 294752). This work is part of the Academy of Finland Flagship Programme, Photonics Research and Innovation (PREIN), decision 320168. It was also funded through the European Metrology Programme for Innovation and Research (EMPIR) project "16ENV09–MetroDecom 2". EMPIR is cofinanced by the Participating States and from the European Union's Horizon 2020 research and innovation program. JK and TT acknowledge financial support from the Jenny and Antti Wihuri Foundation and CHEMS doctoral program of the University of Helsinki, respectively.

This paper is dedicated to the memory of our colleague Dr. Leonid Khriachtchev.